\documentclass[aps,prl,twocolumn,twoside,floatfix,amsmath,showpacs,superscriptaddress]{revtex4-1}
\usepackage{amssymb}
\usepackage{graphicx}
\usepackage{natbib}
\usepackage{color}
\usepackage[normalem]{ulem}
\usepackage{longtable}
\usepackage[utf8]{inputenc}
\begin{document}
%
\newcommand{\CP}{CePd}
\newcommand{\YNS}{YbNiSn}
\newcommand{\CAS}{CeAgSb$_2$}
\newcommand{\YMSG}{Yb$_{14}$MnSb$_{11}$}
\newcommand{\CRPO}{CeRuPO}
\newcommand{\YRCS}{Yb(Rh$_{0.73}$Co$_{0.27}$)$_{2}$Si$_2$}
\newcommand{\YNP}{YbNi$_4$P$_2$}
\newcommand{\YRS}{YbRh$_{2}$Si$_2$}
\newcommand{\betaCNS}{$\beta-$CeNiSb$_{3}$}
\newcommand{\CNS}{CeNiSb$_{2}$}
\newcommand{\YRSb}{YbRhSb}
\newcommand{\YPtG}{YbPtGe}
\newcommand{\YPdG}{YbPdGe}
\newcommand{\CTG}{CeTiGe$_{3}$}
\newcommand{\CIG}{CeIrGe$_{3}$}
\newcommand{\YCS}{YbCu$_{2}$Si$_{2}$}
\newcommand{\CRAB}{CeRu$_{2}$Al$_{2}$B}
\newcommand{\YPS}{YbPdSi}
\newcommand{\CPI}{CePdIn$_{2}$}
\newcommand{\CCG}{CeCrGe$_{3}$}
\newcommand{\CRB}{CeRh$_{3}$B$_{2}$}
\newcommand{\YIG}{YbIr$_3$Ge$_7$}
\newcommand{\CSx}{CeSi$_x$}
\newcommand{\CCAS}{CeCu$_{0.18}$Al$_{0.24}$Si$_{1.58}$}
\newcommand{\CFAPO}{CeFeAs$_{0.7}$P$_{0.3}$O}
\newcommand{\CNP}{CeNi$_{0.8}$Pt$_{0.2}$}
\newcommand{\YRSC}{Yb(Rh$_{1-x}$Co$_{x}$)$_{2}$Si$_{2}$}
%
\newcommand{\sovert}[1]{\ensuremath{#1\,\mu\textnormal{V K}^{-2}}}
\newcommand{\resist}[1]{\ensuremath{#1\,\mu\Omega\textnormal{cm}}}
\newcommand{\mJmolK}[1]{\ensuremath{#1\,\textnormal{mJ mol}^{-1} 
\textnormal{K}^{-2}}}
\newcommand{\lorenzunits}{\ensuremath{\,\textnormal{W \Omega K}^{-2}}}
%
\newcommand{\kB}{\ensuremath{k_{\textnormal{B}}}}
\newcommand{\gfac}{\ensuremath{g_{\textnormal{eff}}}}
\newcommand{\muB}{\ensuremath{\mu_{\textnormal{B}}}}
\newcommand{\TC}{\ensuremath{T_{\textnormal{C}}}}
\newcommand{\TK}{\ensuremath{T_{\textnormal{K}}}}
\newcommand{\TN}{\ensuremath{T_{\textnormal{N}}}}
\newcommand{\To}{\ensuremath{T_{\textnormal{0}}}}
%
\newcommand{\ie}{{\em i.e.}}
\newcommand{\eg}{{\em e.g.}}
\newcommand{\etal}{{\em et al.}}
\newcommand{\replace}[2]{\sout{#1} \textcolor{red}{#2}}
\newcommand{\tred}[1]{\textcolor{red}{#1}}
\newcommand{\tblue}[1]{\textcolor{blue}{#1}}
\title{Kondo-lattice ferromagnets and their peculiar order along the magnetically hard axis}
\author{D.~Hafner}
\affiliation{Max Planck Institute for Chemical Physics of Solids, D-01187 Dresden, Germany}
\author{Binod~K.~Rai}
\affiliation{Department of Physics and Astronomy, Rice University, Houston, Texas 77005, USA}
\author{J.~Banda}
\affiliation{Max Planck Institute for Chemical Physics of Solids, D-01187 Dresden, Germany}
\author{K.~Kliemt}
\author{C.~Krellner}
\affiliation{Physikalisches Institut, Johann Wolfgang Goethe-Universit\"{a}t, D-60438 Frankfurt am Main, Germany}
\author{J.~Sichelschmidt}
\affiliation{Max Planck Institute for Chemical Physics of Solids, D-01187 Dresden, Germany}
\author{E.~Morosan}
\affiliation{Department of Physics and Astronomy, Rice University, Houston, Texas 77005, USA}
\author{C.~Geibel}
\affiliation{Max Planck Institute for Chemical Physics of Solids, D-01187 Dresden, Germany}
\author{M.~Brando}
\affiliation{Max Planck Institute for Chemical Physics of Solids, D-01187 Dresden, Germany}
\date{\today}
%
\begin{abstract}
We show that Ce- and Yb-based Kondo-lattice ferromagnets order mainly along the magnetically hard direction of the ground state Kramers doublet determined by crystalline electric field (CEF). Here we argue that this peculiar phenomenon, that was believed to be rare, is instead the standard case. Moreover, it seems to be independent on the Curie temperature \TC, crystalline structure, size of the ordered moment and type of ground state wave function. On the other hand, all these systems show the Kondo coherence maximum in the temperature dependence of the resistivity just above \TC\ which indicates a Kondo temperature of a few Kelvin. An important role of fluctuations is indicated by the non-mean-field like transition in specific heat measurements as well as by the suppression of this effect by a strong Ising-like anisotropy. We discuss possible theoretical scenarios.
\end{abstract}
\maketitle
Kondo-lattice (KL) systems are typically intermetallic compounds based on trivalent Ce or Yb atoms and are characterized by the Kondo effect at low temperatures and subsequent Kondo coherence at even lower temperatures. The degenerate ground state multiplet ($J = 5/2$ for Ce and $J = 7/2$ for Yb) is split by the crystalline electric field (CEF), making Kramers doublets the prevalent ground state. Only in cubic structures the ground state can be a quartet, which is prone to multipolar order~\cite{Shiina_et_al_1997}. The first excited state is usually located at several tens of Kelvins above the ground state and does not participate in the magnetic ordering. In fact, depending on the strength of the Kondo and Ruderman-Kittel-Kasuya-Yosida (RKKY) interactions, transition temperatures are usually in the order of a few Kelvin, often between 2 and 12\,K, or below 1\,K in systems with a very large distance ($> 6$\,\AA) between the Ce atoms, like in Ce$_{4}$Pt$_{12}$Sn$_{25}$~\cite{Kurita2010}, or strong Kondo effect, like in \YRS~\cite{Trovarelli_et_al_2000}.

It is well established that among all KL systems, there are more than two hundred that show antiferromagnetic (AFM) order at low temperature, while only very few show ferromagnetic (FM) order~\cite{Stewart_2001,Brando_et_al_2016}. The reason for this difference is still unclear. It has recently been proposed that it could result from a $p$-type form factor of the Kondo coupling~\cite{Ahamed_et_al_2018}. To our knowledge, the first FM KL system was discovered by Sato~\etal\ in 1988~\cite{Sato_et_al_1988}. In the following 30 years an increasing number of such compounds have been discovered and studied. The main interest and hope was to find exotic heavy-fermion (HF) superconductivity near a FM quantum critical point (QCP) as it has often been found in the HF AFM systems~\cite{Steglich_et_al_1979,Mathur_et_al_1998}. But up to now, apart from some U-based ferromagnets~\cite{Aoki_et_al_2011}, no Ce- or Yb-based FM KL superconductor has been found. One of the reasons for the absence of superconductivity in quantum critical metallic ferromagnets has been suggested to be the absence of quantum critical fluctuations due to the intrinsic first order phase transition at the FM quantum phase transition~\cite{Belitz_et_al_1999,Brando_et_al_2016}.

The discovery of \YNP~\cite{Krellner_et_al_2011} with the lowest Curie temperature among pure compounds ever observed (\TC\ = 0.15\,K), and the subsequent observation of a FM QCP \cite{Steppke_et_al_2013}, reopened the discussion about the existence of FM QCPs~\cite{Steppke_et_al_2013,Gegenwart_et_al_2015}. Along with this unexpected observation, another peculiar feature in this system caught our attention: The magnetic moments in the FM ordered state point along the magnetically CEF hard axis and not, as naively expected, along the easy axis~\cite{Comment_01}. In fact, ordering along the easy direction (with a larger moment) is expected because the gain in energy in the ordered state is proportional to the square of the size of the ordered moment. Such a behavior was first observed by Bonville~\etal\ in \YNS\ with a \TC\ of 5.6\,K~\cite{Bonville_et_al_1992} and it was surprisingly also found in \YRCS, a tetragonal system with a large CEF magnetic anisotropy of about 6~\cite{Lausberg_et_al_2013,Hamann_et_al_2018}. There is no clear explanation for this counter-intuitive phenomenon, but at least two theoretical approaches were recently proposed: i) A Monte-Carlo calculation based on the Heisenberg model with competing FM and AFM ordering combined with competing anisotropies in exchange interactions and $g$-factors, which could reproduce the experimental results for \YRCS\ very well~\cite{Andrade_et_al_2014}. This model, however, seems not to work for large CEF anisotropies~\cite{Hamann_et_al_2018} and does not provide any explanation for the particular choice of exchange couplings in the system, and it does not work for quantum critical systems like \YNP. ii) Another and more general approach is the one proposed by Kr\"uger~\etal: Large fluctuations in an easy basal plane favor ordering along the transversal hard axis~\cite{Kruger2014}. The idea of having a fluctuations induced transition would work for any classical or quantum ferromagnet, provided that the anisotropy is not too large.

In the course of our studies on other KL ferromagnets like \CRPO~\cite{Krellner_Geibel_2008} and \YIG~\cite{Rai_et_al_2019}, we have realized that even more of these systems show this peculiar behavior. Here, we present part of these studies and compile a list of FM KL systems to show that this phenomenon, rather than a rare occurrence, is instead the general case.
\begin{figure}[t]
	\begin{center}
		\includegraphics[width=\linewidth]{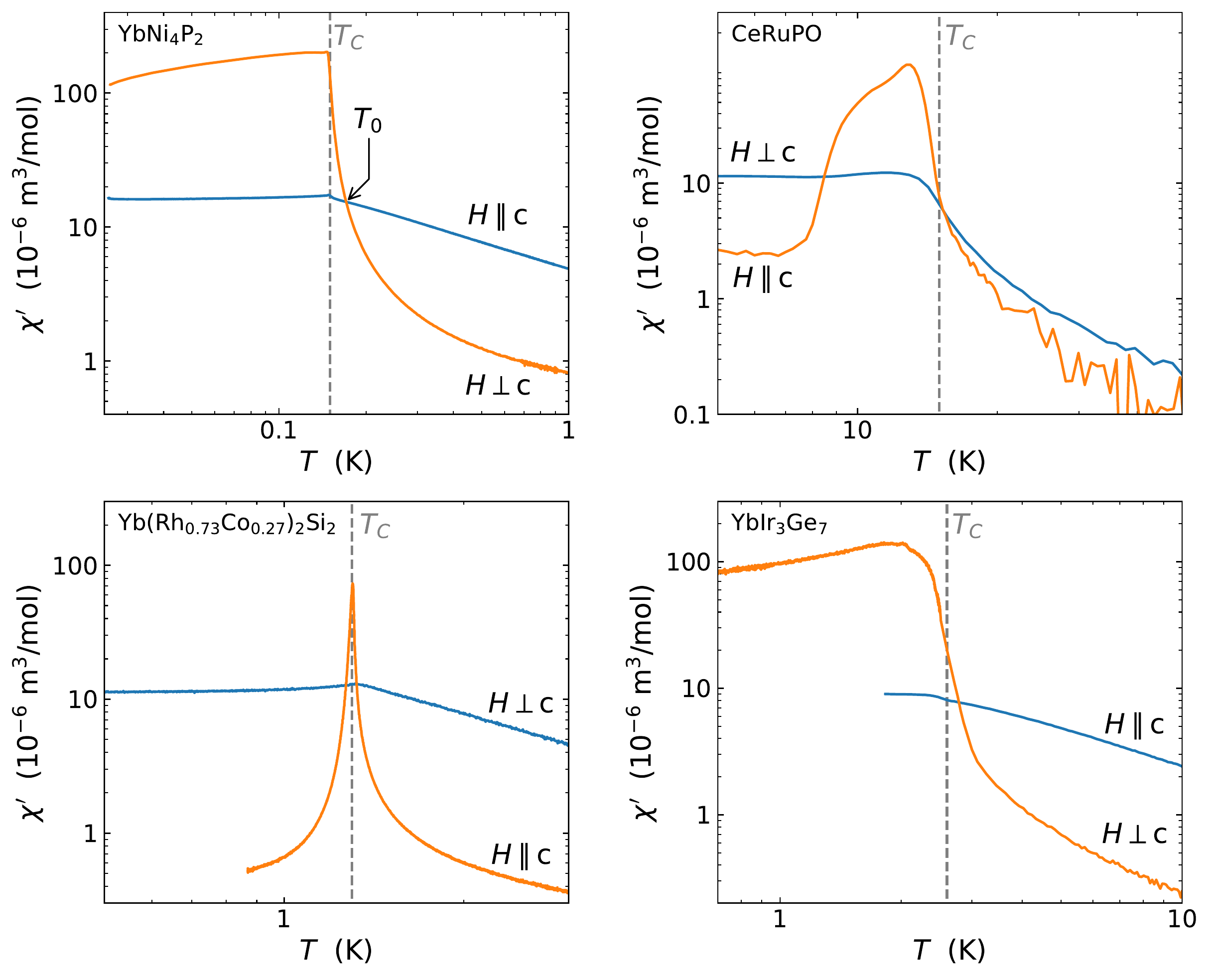}
	\end{center}
	\caption{Temperature dependence of the magnetic ac-susceptibility $\chi'(T)$ for \YNP, \YRCS, \CRPO\ and \YIG\ measured with modulated field along the two principal directions ($H_{ac} \parallel c$ and $H_{ac} \perp c$) of the tetragonal crystalline structure. \YNP, \YRCS\ graphs are reproduced from ~\cite{Steppke_et_al_2013},\cite{Lausberg_et_al_2013} respectively. Strong noise is seen for \CRPO\ due to the very small size of the crystal.}
	\label{fig1}
\end{figure}

We start showing the main properties of these FM KL systems that order along the hard axis by taking a look at the temperature dependence of the magnetic ac-susceptibility  $\chi'(T)$ of \YNP, \YRCS, \CRPO, and \YIG\ plotted in Fig.~\ref{fig1}. $\chi'(T)$ was measured with modulated field along the two principal directions of the tetragonal crystalline structure, $H_{ac} \parallel c$ and $H_{ac} \perp c$. At high temperatures, both susceptibilities follow the same $T$-dependence, because of the dominant Curie-Weiss contribution of the full moment of trivalent Ce and Yb. However, their absolute values differ significantly due to the magnetocrystalline anisotropy caused by the CEF of the tetragonal structure. At low temperatures, just above \TC , these susceptibilities cross each other at a temperature $\To\ $ (marked by an arrow in the first panel of Fig.~\ref{fig1}), which inevitably indicates that the magnetic moments order along the magnetically hard direction of the CEF. This has also been confirmed by magnetization measurements at $T < \TC$~\cite{Steppke_et_al_2013,Lausberg_et_al_2013,Krellner_Geibel_2008,Rai_et_al_2019}. Below \TC, the measured $\chi'(T)$ perpendicular to the ordered moments remains constant, while the behavior of $\chi'(T)$ parallel to the moments depends on the ratio between the coercive field and the modulated field used in the measurements: For instance, in \YNP\ $\chi'(T)$ stays constant below \TC\ whereas in \YRCS\ it decreases steeply. The fact that \To\ is just above \TC\ implies that there is no correlation between this behavior and the CEF first excited state which is located at much higher temperatures in these systems. There are indeed FM systems in which the susceptibilities cross each other at a high \To\ because of the CEF excited states like \CSx~\cite{Sato_et_al_1988,Pierre_et_al_1990} (\To\ $\approx 70$\,K) or \CCAS~\cite{Maurya_et_al_2017} (\To\ $\approx 40$\,K).

To demonstrate that this peculiar behavior, that was believed to be rare, is instead the general rule, we present a list of all KL ferromagnets known to date in table~\ref{table}. There are twelve systems that order along the hard axis and only two exceptions, \CTG~\cite{Fritsch_et_al_2015} and \CRAB~\cite{Matsuoka_et_al_2012,Baumbach_et_al_2012}. We also present in the following other measurements performed on the same four systems of Fig.~\ref{fig1} to emphasize some characteristic properties that are common to all KL ferromagnets, which will help us to derive some conclusions about the origin of this behavior. We should also mention that our analysis is valid for systems with a single site for Ce or Yb. Systems with more sites for Ce or Yb are obviously more complex and might deviate from the general trend, like in the case of \YPdG~\cite{Enoki_et_al_2012,Tsujii_et_al_2015}. However, \YPS~\cite{Tsujii_et_al_2016} and \betaCNS~\cite{Thomas_et_al_2007} have more than one site per magnetic atom, but order along the hard axis.
\begin{figure}[t]
	\begin{center}
		\includegraphics[width=\linewidth]{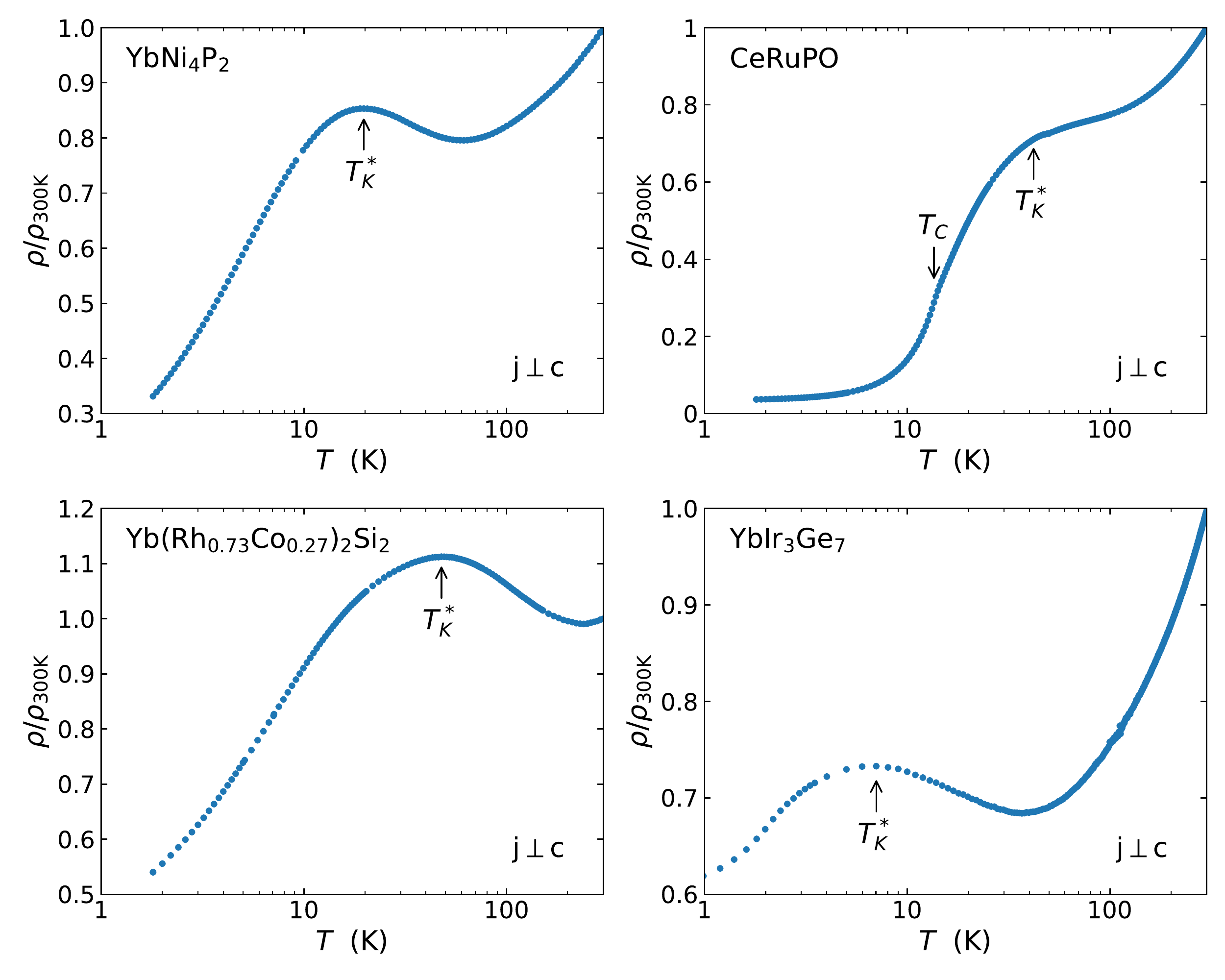}
	\end{center}
	\caption{Temperature dependence of the resistivity $\rho$ divided by the room temperature resistivity $\rho_{\textnormal{300K}}$ plotted in a logarithmic scale to emphasize the maximum at $T_{K}^{*}$ in all four materials which is due to the Kondo coherence effect. \YNP, \CRPO, \YRCS, \YIG\ graphs are reproduced from ~\cite{Kliemt_Krellner_2017},\cite{Krellner_Geibel_2008},\cite{Klingner_et_al_2011},\cite{Rai_et_al_2019} respectively.}
	\label{fig2}
\end{figure}
\begingroup
\squeezetable
\begin{table*}[t]
	\caption{Ce- and Yb-based Kondo-lattice ferromagnets. YRCS = \YRCS, \TC = Curie temperature, \TK = Kondo temperature, $\mu$ = magnetic moment, CEF = crystalline electric field, ESR = electron spin resonance (9.4\,GHz, $T>3$\,K, main crystalline directions), MFT = mean-field transition, GS = ground state wave function, n.a. = not available, $^{*}$ = moments order along the CEF magnetically hard axis.}
	\smallskip
	\begin{ruledtabular}
		\begin{tabular}{lcclcclllccl}
			System 
			& \TC\,(K) 
			& \TK\,(K)\,$^{a}$ 
			& crys. 
			& ordered
			& CEF
			& easy
			& order
			& ESR
			& MFT 
			& coherence\\
			&
			&                           
			& struc.
			& $\mu$ ($\mu_{\text{B}}$)\,$^b$
			& anisotropy
			& axis
			& axis
			& signal
			& 
			& max. in  $\rho(T)$\\
			\hline\\[-7pt]
			$^{*}$\CAS~\cite{Myers_et_al_1999,Andre_et_al_2000,Sidorov_et_al_2003,Araki_et_al_2003,Takeuchi_et_al_2003, Jobiliong_et_al_2005}
			& 9.6
			& 16
			& tetra. 
			& 0.41
			& 3
			& $ab$-plane
			& $c$-axis
			& not found
			& weak
			& yes\\
			\\[-5pt]
			$^{*}$\CRPO~\cite{Krellner_et_al_2007,Krellner_Geibel_2008,Krellner_et_al_2008}
			& 15
			& 7
			& tetra. 
			& 0.3
			& 3
			& $ab$-plane
			& $c$-axis
			& found
			& no
			& yes\\
			\\[-5pt]
			$^{*}$\CFAPO~\cite{Jesche_et_al_2012}
			& 7.5
			& 5
			& tetra. 
			& 0.3
			& 3
			& $ab$-plane
			& $c$-axis
			& found
			& no
			& yes\\
			\\[-5pt]
			$^{*}$YRCS~\cite{Klingner_et_al_2011,Lausberg_et_al_2013,Gruner_et_al_2012}
			& 1.3
			& 7.5
			& tetra. 
			& 0.1
			& 6
			& $ab$-plane
			& $c$-axis
			& found
			& no
			& yes\\
			\\[-5pt]
			$^{*}$\YNP~\cite{Krellner_et_al_2011,Krellner_Geibel_2012,Steppke_et_al_2013,Gegenwart_et_al_2015}\,$^{b}$
			& 0.15
			& 8
			& tetra.
			& $< $ 0.05
			& 5
			& $c$-axis
			& $ab$-plane
			& not found
			& no
			& yes\\
			\\[-5pt]
			$^{*}$\YIG~\cite{Rai_et_al_2019}
			& 2.6
			& 16
			& rhomb. 
			& 0.05
			& 4
			& $c$-axis
			& $ab$-plane
			& not found
			& no
			& yes\\
			\\[-5pt]
			$^{*}$\YNS~\cite{Bonville_et_al_1992,Kasaya_et_al_1991,Generalov_et_al_2017}
			& 5.6
			& 10
			& orth. 
			& 0.85
			& 1.8
			& $a$-axis
			& $c$-axis
			& not found 
			& no
			& yes\\
			\\[-5pt]
			$^{*}$\YPtG~\cite{Katoh_et_al_2009}
			& 5.4
			& 9.4
			& orth. 
			& 1
			& 2
			& $a$-axis
			& $c$-axis
			& n.a.
			& no 
			& yes\\
			\\[-5pt]
			$^{*}$\YRSb~\cite{Muro_et_al_2004,Umeo_et_al_2012}\,$^{c}$
			& 4.3
			& 30
			& orth. 
			& 0.4
			& 2
			& $a$-axis
			& $c$-axis
			& n.a.
			& no
			& yes\\
			\\[-5pt]
			$^{*}$\YPS~\cite{Tsujii_et_al_2016}\,$^{d}$
			& 8
			& 13
			& orth.
			& 0.26
			& n.a.
			& $b$-axis
			& $c$-axis
			& n.a.
			& no
			& yes\\
			\\[-5pt]
			$^{*}$\betaCNS~\cite{Thomas_et_al_2007}\,$^{e}$
			& 6
			& 10
			& orth.
			& 0.9
			& 1.5
			& $b$-axis
			& $c$-axis
			& n.a.
			& no
			& yes\\
			\\[-5pt]
			$^{*}$\CIG~\cite{Muro_et_al_1998,Kawai_et_al_2008,Anand_et_al_2018}\,$^{f}$
			& 4.8
			& 12
			& tetra. 
			& 0.14
			& 1.3
			& $a$-axis
			& $c$-axis
			& n.a.
			& 1$^{st}$-order
			& yes\\
			\\[-5pt]
			\hline\\[-7pt]
			\CTG~\cite{Fritsch_et_al_2015}
			& 14
			& 30
			& hexag. 
			& 1.5
			& 10
			& $c$-axis
			& $c$-axis
			& n.a.
			& yes
			& yes\\
			\\[-5pt]
			\CRAB~\cite{Matsuoka_et_al_2012,Baumbach_et_al_2012,Bhattacharyya_et_al_2016}\,\,$^g$
			& 13
			& 23
			& tetra. 
			& 1
			& $> 40$
			& $c$-axis
			& $c$-axis
			& n.a.
			& n.a.
			& yes\\
			\\[-10pt]
			\hline\hline\\[-5pt]
			\multicolumn{12}{l} {$^a$ Reported or estimated from entropy.\quad 
				$^b$ In-plane anisotropy: orthorhombic point symmetry site for Yb.}\\
			\multicolumn{12}{l} {$^{c}$ Under pressure of about 2\,GPa. At zero pressure the order is canted AFM with a very small ordered moment (0.003\,\muB\,}\\
			\multicolumn{12}{l} {along the $b$-axis). \TK\ is the value at zero pressure.}\\
			\multicolumn{12}{l} {$^d$ Largest moment in a complex structure with 3 Yb sites and 3 different moment sizes. $^e$ Two Ce sites.}\\
			\multicolumn{12}{l} {$^{f}$ Transition into a canted AFM at 8.7\,K, which is probably 1$^{st}$-order. Recent neutron experiments suggest the magnetic}\\
			\multicolumn{12}{l} {structure is more complex than a collinear FM~\cite{Anand_et_al_2018}}\\
			\multicolumn{12}{l} {$^g$ AFM transition at 14.3\,K.}\\
		\end{tabular}
	\end{ruledtabular}
	\vskip -3mm
	\label{table}
\end{table*}
\endgroup
\begin{figure}[t]
	\begin{center}
		\includegraphics[width=\linewidth]{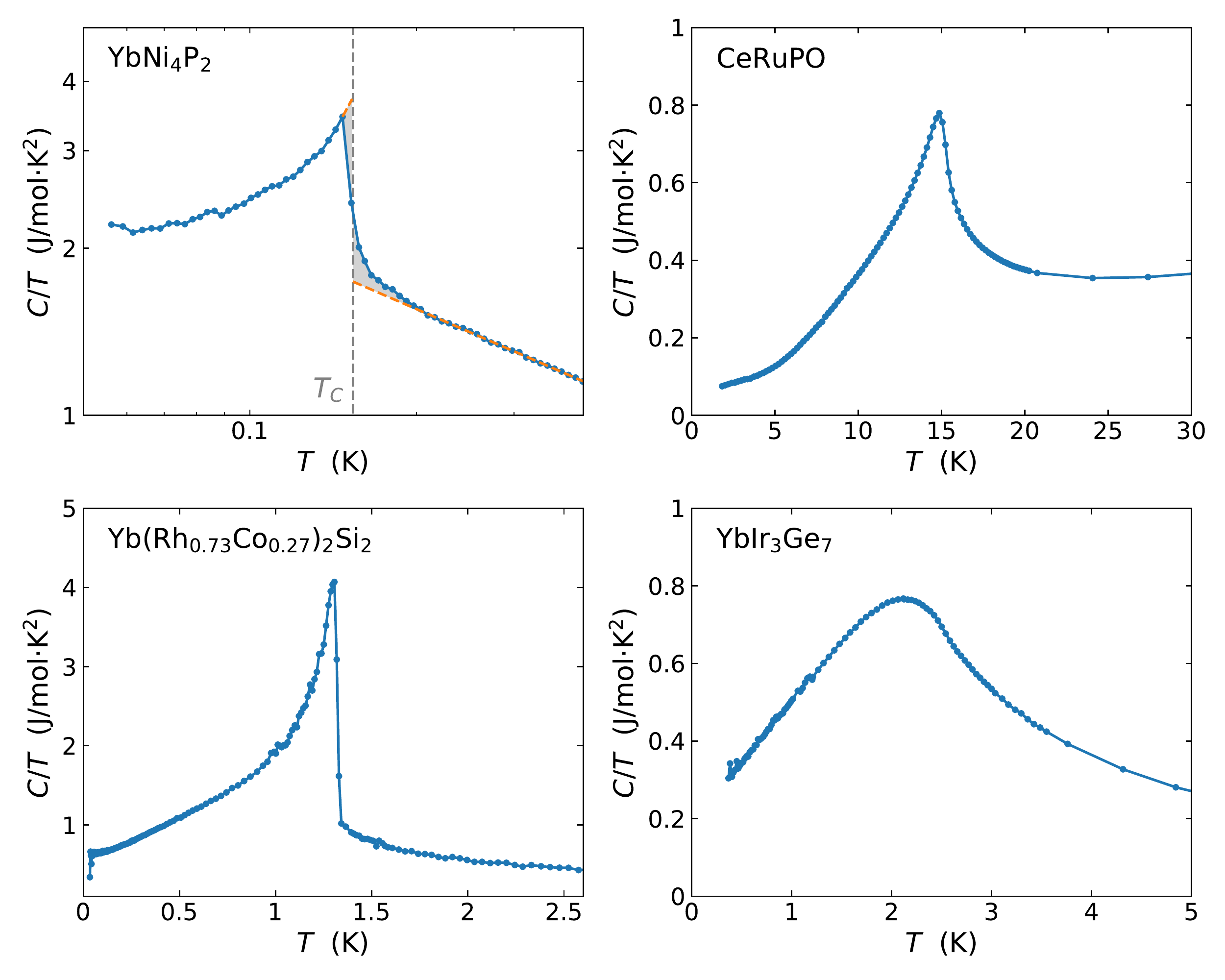}
	\end{center}
	\caption{Temperature dependence of the specific heat around the FM phase transition. Transition temperatures are listed in Table~\ref{table}. To emphasize the presence of critical fluctuations of the ferromagnetic transition, the \YNP\ data are exemplarily plotted in a logarithmic scale and compared to a MFT. \YNP , \CRPO, \YRCS, \YIG\ graphs are reproduced from ~\cite{Steppke_et_al_2013},\cite{Krellner_et_al_2007},\cite{Lausberg_et_al_2013},\cite{Rai_et_al_2019} respectively.}
	\label{fig3}
\end{figure}

A common property of KL ferromagnets is the presence of the Kondo effect with a Kondo temperature $T_{K}$ of a few Kelvins, often close to \TC. $T_{K}$ is listed in table~\ref{table} and, if not reported, it was estimated by us from the magnetic entropy $S_\textnormal{m}$ extracted from specific heat measurements with $S_\textnormal{m}(\frac{1}{2}T_\textnormal{K})=\frac{1}{2}R\ln 2$. The presence of the Kondo effect can also be seen in the temperature dependence of the resistivity, exemplarily plotted in Fig.~\ref{fig2} for the same four systems of Fig.~\ref{fig1}. The maximum in $\rho(T)$ at $T_{K}^{*} > \TC$\ indicates the Kondo coherence temperature~\cite{Krellner_et_al_2011,Krellner_et_al_2007,Klingner_et_al_2011,Rai_et_al_2019}. This behavior is similar in all ferromagnets showing order along the hard axis. In contrast to this, materials in which the Kondo effect is absent or $T_{K}$ is small show ordering along the easy axis, like \CNS~\cite{Thamizhavel_et_al_2003}. Although \CTG~\cite{Fritsch_et_al_2015} and \CRAB~\cite{Matsuoka_et_al_2012,Baumbach_et_al_2012} show a sizeable $T_{K}^{*}$, they order along the easy axis. A hint why these two systems do not follow the common rule can be found in their huge Ising-type anisotropy, which limits fluctuations to longitudinal ones along the c-axis. This would strongly reduce the possibility for fluctuation induced order. This assumption is also supported by the shapes of the second order phase transition at \TC\ measured in specific heat. While for \CTG\ the transition is mean-field-like as expected for an Ising-system, for the other compounds ordering along the hard axis, the transistion is $\lambda$-shaped, signifying that strong fluctuations are present around \TC. This is shown in Fig.~\ref{fig3} for the same four compounds of Figs.~\ref{fig1} and \ref{fig2}. 

Regarding theoretical proposals, the evidence for fluctuations in the temperature dependence of the specific heat near \TC\ also strongly favors the theory by Kr\"uger \etal~\cite{Kruger2014} based on strong transversal fluctuations over a purely MFT as the one based on competing anisotropies by Andrade \etal~\cite{Andrade_et_al_2014}. In addition, it has recently been shown that in \YRSC\ a model with competing anisotropy of the exchange interaction which is supposed to overcome the CEF anisotropy is unlikely, because the huge CEF anisotropy ($> 10$) for small $x$ would require a huge inverse anisotropy in the exchange interactions~\cite{Hamann_et_al_2018}.
A direct approach to get information on exchange interactions is to measure the dispersion relation of magnetic excitation, e.g. magnons, using inelastic neutron scattering (INS). Therefore one might expect that the anisotropy of the exchange interactions in the FM systems is a problem which can unambiguously be settled using this approach. For one of the compounds discussed here, CeAgSb$_2$, such a detailed INS study has been performed~\cite{Araki_et_al_2003}. This study indeed concluded that all experimental results, including the FM ordering along the hard axis, can be fully explained by a huge anisotropy of the exchange interactions~\cite{Araki_et_al_2003}. As we show in detail in the supplemental material, however, there are problems in their analysis: In contrast to the view suggested in \cite{Araki_et_al_2003}, the magnon dispersion relations typically determined by INS do not provide a unique answer concerning exchange interactions on their own. Specifically in the case of CeAgSb$_2$, we also demonstrate that the huge anisotropy of the exchange parameters deduced from INS in \cite{Araki_et_al_2003} are in clear contradiction to the anisotropy of the susceptibility observed at high temperatures. This contradiction points on its own to a yet unidentified phenomena which promotes the hard axis ordering at low temperature. A similar discrepancy between a huge anisotropy of exchange interactions deduced from INS experiments and a much weaker anisotropy of the susceptibility at high temperature is also observed in the system CeRu$_2$Al$_{10}$ with AFM order along the hard axis~\cite{Robert_et_al_2012, Strigari_et_al_2012, Dean_et_al_2018, Adroja_2013}. The authors of \cite{Robert_et_al_2012} conducted this type of analysis and concluded that this leads to an "unrealistically large" value for the hard axis exchange interaction. The similarity between these cases suggests the problem addressed in our paper is not only relevant to FM, but also for AFM Kondo systems.

Table~\ref{table} also shows what these KL ferromagnets do not have in common: For instance, the crystalline structure, the size of the ordered moment or \TC. The ground state wave functions are also very different. While it has been proposed that FM correlations are essential for the observability of an ESR signal in a KL system~\cite{Abrahams_2008,Kochelaev_2009}, the situation still seems unclear. Inspired by a work done on some KL systems a few years ago~\cite{Krellner_et_al_2008}, in which the detection of the ESR signal at 9.4\,GHz was attributed to the presence of FM correlations, we investigated the ESR response of some of these systems and included whether such a signal has been found or not in Tab.~\ref{table}. Unfortunately, there does not seem to be any systematic relation and although the ESR signal is undoubtedly connected to ferromagnetism, the latter is not the only deciding factor for the occurrence of an ESR signal.

Having ruled out the role of CEF excited states and a model with competing exchange interactions, and considering that the only common features between the systems in Tab.~\ref{table} are the presence of the Kondo effect, fluctuations at \TC\ and the possibility of sizeable (transversal) fluctuations perpendicular to the hard directions, it seems that the most possible scenario is that in which fluctuations are the driving force, a sort of order-by-disorder mechanism like the one proposed by Kr\"uger \etal~\cite{Kruger2014}. However, while there is a qualitative match for the susceptibility curves between theory and experiment, there are still other details that do not match, e.g. the proposed first order transition versus the observed second order transition. Further comparisons require detailed measurements of the direction dependence of fluctuations, which are possible by neutron scattering or NMR experiments. More systems and information might also be needed to finally unravel the origin of this mysterious behavior. In fact, there are some FM systems for which only polycrystals are available, like CePd~\cite{Kappler_et_al_1991} or \CPI~\cite{daSilva_et_al_2009,Carleschi_et_al_2015}; or systems which show FM ordering only at very high pressure, like \YCS~\cite{Tateiwa_et_al_2014,Dung_et_al_2009} (at about 11.5\,GPa), for which not much information about the CEF anisotropy at high-$p$ is available. We would also like to mention, that this phenomenon has also been observed in cerium and actinide intermetallics~\cite{Cooper_Siemann_1979} and some AFM KL systems, i.e. CeRu$_{2}$Al$_{10}$~\cite{Kondo_et_al_2013},  CeOs$_{2}$Al$_{10}$~\cite{Khalyavin_et_al_2013} and CeRhIn$_5$~\cite{Takeuchi_2001}. In these systems the change from hard-axis to the easy-axis ordering has been attributed to the weakening of the Kondo hybridization. But this seems to be in contrast to what has been observed, e.g., in \YRSC, in which increasing the Kondo hybridization drives the moments into the easy plane~\cite{Klingner_et_al_2011,Hamann_et_al_2018}.

In conclusion, we observe that Ce- and Yb-based Kondo-lattice ferromagnets order mainly along the magnetically CEF hard direction. This behavior is independent on \TC, crystalline structure, size of the ordered moment and type of ground state wave function. On the other hand, all these systems show Kondo temperatures of a few Kelvin, often close to \TC, and they have in common a relatively small CEF anisotropy. CEF excited states are too high in energy to be responsible for this behavior. Specific heat measurements indicate that the second order phase transition is not mean-field like, pointing to an important role of fluctuations, which might induce such an order along the hard axis. However, the intrinsic mechanism leading to this kind of order in all KL ferromagnets remains unknown. We further note that a huge Ising-type anisotropy prevents this unexpected type of ordering and leads to conventional order along the easy axis.
\begin{acknowledgments}
We would like to thank D. Adroja, O. Erten, A. Green, L. Havela, D. Kaczorowski, S. Kambe, F. Kr\"uger, P. Riseborough, P. Ronning, J. Spałek and A. Strydom for inspiring discussions and the DFG for financial support from project BR 4110/1-1 and KR 3831/4-1. This research is funded in part by a QuanEmX grant from ICAM and the Gordon and Betty Moore Foundation through Grant GBMF5305 to B.~K.~Rai. E.~Morosan acknowledges partial support from the Gordon and Betty Moore Foundation EPiQS' initiative through grant GBMF4417, and travel support to the MPI in Dresden, Germany from the Alexander von Humboldt Foundation Fellowship for Experienced Researchers.
\end{acknowledgments}

\bibliography{KLF_PRL_hafner}

\begin{thebibliography}{10}
\bibitem{Comment_01}
Here we define the magnetically easy axis as the crystallographic axis of easy
  magnetization determined by the CEF.
\end{thebibliography}
\bibliographystyle{h-physrev}

\end{document}